\documentclass[amsmath,amssymb,jcp,twocolumn,superscriptaddress]{revtex4-1}

\begin{document}
\title{
Mixed quantum-classical evolution in open molecular systems.
}

\author{Michael Galperin}
\email{mgalperin@tauex.tau.ac.il}
\affiliation{School of Chemistry, Tel Aviv University, Tel Aviv 6997801, Israel}
\author{Abraham Nitzan}
\email{anitzan@sas.upenn.edu}
\affiliation{Department of Chemistry, University of Pennsylvania, Philadelphia,
Pennsylvania 19104, United States}

\begin{abstract}
We consider a mixed quantum-classical formulation for open nonequilibrium molecular systems. We employ the pseudoparticle nonequilibrium Green's function (PP-NEGF) method as a fully quantum description and a starting point for introducing classical nuclear dynamics for an open system. We use this formulation to derive a density-matrix equation of motion describing the quantum-classical evolution. We compare our results with previous formulations and highlight important differences related to the nonequilibrium and open character of the molecular system.
We also analyze the approximations needed to reduce full quantum-classical dynamics to the fewest-switches surface hopping (FSSH) method and show that these approximations are unreasonable when adiabatic surfaces come close to each other.
\end{abstract}

\maketitle

\section{Introduction}
Nonadiabatic molecular dynamics (NAMD) is central for the theoretical description of many processes, including charge transfer~\cite{asbury_ultrafast_2001,duncan_theoretical_2007,giannini_charge_2022,warburton_theoretical_2022}, energy transfer and dissipation~\cite{meng_dynamics_2026}, solar energy conversion~\cite{long_unravelling_2016,ponseca_ultrafast_2017,zheng_ab_2019,chu_low-frequency_2020}, intersystem crossing~\cite{penfold_spin-vibronic_2018}, and chemical processes~\cite{chu_time-dependent_2006,herrera_molecular_2020,mandal_theoretical_2023}.
Thus, NAMD is naturally the focus of many theoretical studies where nuclear degrees of freedom are treated classically while electron subsystem remains quantum~\cite{ben-nun_ab_2000,worth_beyond_2004,stock_classical_2005,jasper_non-born-oppenheimer_2006,levine_isomerization_2007,yarkony_nonadiabatic_2011,akimov_theoretical_2013,casanova_theoretical_2018,curchod_ab_2018,crespo-otero_recent_2018,ivakhnenko_nonadiabatic_2023,omranpour_machine_2025,zhang_machine_2025}.
The most efficient and widely used in practical simulations is the fewest-switches surface hopping (FSSH) algorithm~\cite{tully_molecular_1990}.
An interesting recent development is the phase-space approach by Subotnik and co-workers~\cite {bian_phase-space_2026}.

The FSSH is an {\em ad hoc} formulation of the NAMD, and much effort has been made to derive and/or improve it~\cite{subotnik_new_2011,nelson_nonadiabatic_2013,ouyang_surface_2015,sifain_mixed_2015,wang_recent_2016,subotnik_understanding_2016,martens_classical_2019,martens_surface_2019,huang_first_2023}.
Most of these studies focused on isolated molecular systems.
Efforts to generalize the formulation to open systems can be divided into a brute-force approach (consideration of system and bath as one isolated supersystem)~\cite{roy_dynamics_2009,shenvi_nonadiabatic_2009} and more elaborate approaches~\cite{ouyang_surface_2015,dou_surface2_2015,dou_surface3_2015}.
Still, the latter studies partly rely on plausible approximations and do not provide a consistent first-principles derivation. 
An attempt at a consistent approach to the FSSH was made by considering a first-principles derivation of quantum-classical equations~\cite{kapral_mixed_1999,kapral_quantum-classical_2001,sergi_quantum-classical_2004,kim_transport_2005,kapral_progress_2006,grunwald_decoherence_2007,hsieh_analysis_2013} with subsequent introduction of an approximation to reduce them to the FSSH formulation~\cite{kapral_surface_2016}.
This research was mostly focused on isolated molecules.
An approach to open systems in Ref.~\cite{kapral_progress_2006} extended the isolated-system consideration by separating classical and quantum dynamics, followed by tracing out parts of the (initially isolated) supersystem. 
Limitations of this approach are, first, the inability to consider nonequilibrium situations with several baths (each at its own equilibrium) coupled to the molecule and, second, 
the ambiguity regarding the order of operation: transfer to classicality of the nuclear subspace dynamics and tracing out bath degrees of freedom.
Making the classical approximation before tracing out the thermal environment partially disregards quantum effects at the molecule-bath interface and may lead to a wrong equilibrium distribution at long times. More generally, traditional treatments of such open systems, using methods originally developed to describe isolated molecules, are minimally inconsistent and maximally questionable. For example, such are theoretical descriptions of photo- and bias-induced chemistry for molecules on surfaces and at interfaces. Indeed, the very concepts of adiabatic surface, conical intersection, and so on are defined only for an isolated molecule. 

Here, we generalize the mixed quantum-classical evolution
originally developed in Refs.~\cite{kapral_mixed_1999,kapral_progress_2006}
to open molecular systems (e.g., single-molecule junctions). We start from a fully quantum description and trace out bath degrees of freedom before considering the quantum-to-classical transition. 
We extend our previous study~\cite{galperin_nuclear_2015} to include both Fermi (contacts) and Bose (thermal) baths, compare our approach with earlier quantum-classical formulations, and analyze the approximations necessary to arrive at the FSSH formulation.

The structure of the paper is as follows. Section~\ref{ppnegf} presents a fully quantum formulation for the open molecular system employing the pseudoparticle nonequilibrium Green's function (PP-NEGF). Section~\ref {BO_ppnegf} discusses the quantum-to-classical transition in nuclear dynamics within the PP-NEGF formulation. Section~\ref{dm} introduces a generalized (energy-resolved) version of the density matrix and uses the previously discussed PP-NEGF formulation to derive its equation of motion. We compare the formulation with previous quantum-classical considerations. In Section~\ref{SectionFSSH}, we analyze the approximations needed to reduce the formulation to the FSSH consideration. Section~\ref{conclusion} summarizes our findings and indicates future research directions.


\section{PP-NEGF formulation}\label{ppnegf}
We start from the first-quantization form of a Hamiltonian 
representing a molecule $M$ with $N_n$ nuclei and $n_e$ electrons.
In addition, the molecular nuclei are coupled to environmental degrees of freedom modeled as a thermal bath of harmonic oscillators.
Similarly, the electronic subsystem encompasses the molecule together with substrate (electrode) degrees of freedom.
 For simplicity and following Ref.~\cite{kapral_mixed_1999} we rperesent all nuclear 
degrees of coordinates with single operator $\hat Q\equiv (\hat Q_1,\ldots,\hat
Q_{N})$, all electron coordinates are designated with
$\hat q\equiv(\hat q_1,\ldots,\hat q_{n})$. 
We use similar notation for momentum operators.
$M$ and $m$ are, respectively,
masses of nuclei and electrons (again, for nuclei $M$ may designate $N$
different values). 
An explicit expression for the Hamiltonian is
\begin{equation}
\label{H}
\hat H = \frac{\hat P^2}{2M} + \frac{\hat p^2}{2m} + V(\hat q,\hat Q)
\end{equation}
Below, we will use the basis of eigenstates of operators $\hat Q$ and $\hat q$
where (here and below $\hbar=1$)
\begin{equation}
\begin{split}
  \langle Q\rvert \hat P \lvert Q'\rangle
  &= -i\delta(Q-Q')\frac{\partial}{\partial Q}
  \\
  \langle q\rvert \hat p \lvert q'\rangle
  &= -i\delta(q-q')\frac{\partial}{\partial q}
\end{split}
\end{equation}

Our goal is the evaluation of the Green's function, whose knowledge allows
to simulate multiple system responses under external perturbations.
Because molecular dynamics is usually described within a many-body basis
of molecular states, we employ a many-body flavor of the nonequilibrium
Green's function (NEGF) method - the pseudoparticle NEGF (PP-NEGF).
In complete analogy with the standard (single-particle) second quantization
we introduce operator of quantum field $\hat\Psi(q,Q)$ 
which represents the state of the
molecular system at points $q$ (for electrons) and $Q$ (for nuclei) 
and consider a pseudoparticle Green's function defined on the Keldysh contour as
\begin{equation}
\label{G}
  G(q,Q,\tau;q',Q',\tau') \equiv -i\langle T_c\,\hat\Psi(q,Q,\tau)\,
  \hat\Psi^\dagger(q',Q',\tau')\rangle
\end{equation}
Here, $T_c$ is the contour ordering operator, $\tau,\tau'$ are the
contour variables, the quantum fields are in the Heisenberg picture, and the average is over the quantum statistical ensemble.
The Dyson equation describes the evolution of the Green's function
\begin{align}
\label{Dyson}
   &i\frac{\partial}{\partial\tau} G(q,Q,\tau;q',Q',\tau') =
   \\ &
  \left[-\frac{1}{2M}\frac{\partial^2}{\partial Q^2}
  -\frac{1}{2m}\frac{\partial^2}{\partial q^2}+V(q,Q)\right]
  G(q,Q,\tau;q',Q',\tau')
 \nonumber \\ &
    + \int_c d\tau'' \int dq''\int dQ''\,\Sigma(q,Q,\tau;q'',Q'',\tau'')
 \nonumber   \\ & \times
    G(q'',Q'',\tau'';q',Q',\tau')
        +\delta(q-q')\delta(Q-Q')\delta(\tau,\tau')  
  \nonumber
\end{align}
Here,  $\Sigma$ is the self-energy describing the effect of the thermal environment and of contacts on the molecular subsystem.
For the derivation, please see Appendix~\ref{appA}.


\section{Quantum-to-classical transition in nuclear dynamics}\label{BO_ppnegf}
Because nuclei are much heavier than electrons, $M\gg m$, their dynamics can be assumed to be almost classical,
while the electron subsystem still requires a strict quantum-mechanical description.
A standard way of separating quantum and classical dynamics on the contour
is provided by the gradient expansion~\cite{haug_quantum_2008}.

We start from the Dyson equation (\ref{Dyson}) and perform the Wigner transform in nuclear coordinates
\begin{equation}
R\equiv\frac{Q+Q'}{2},\qquad Z\equiv Q-Q'
\end{equation}
followed by the Fourier transform in the quantum coordinate: $Z\to P$.
Assuming weak molecule-bath couplings, we employ a zero-order gradient expansion for which
\begin{equation}
\int dQ''\, A(Q,Q'')\, B(Q'',Q') \approx A(R,P)\, B(R,P) 
\end{equation}
This leads to
\begin{align}
  \label{LDyson_GE}
    & i\frac{\partial}{\partial\tau} G(q,\tau;q',\tau';R,P) =  
    \bigg[
      -\frac{1}{2M}\left(\frac{1}{4}\frac{\partial^2}{\partial R^2}
      +iP\frac{\partial}{\partial R} - P^2\right)
      \nonumber  \\ &
      + \hat h(q,R)
      +\frac{i}{2}\frac{\partial V(q,R)}{\partial R}\frac{\partial}{\partial P}
    \bigg] G(q,\tau;q',\tau';R,P)
  \nonumber \\ &
    +\int_c d\tau''\int dq''\, \Sigma(q,\tau;q'',\tau'';R,P)\,
    G(q'',\tau'';q',\tau';R,P)
  \nonumber \\ &
    +\delta(q-q')\,\delta(\tau,\tau')
\end{align}
where
\begin{equation}
\hat h(q,R) \equiv -\frac{1}{2m}\frac{\partial^2}{\partial q^2} + V(q,R)
\end{equation}
Note that the analogous separation of classical and quantum
motion for an isolated molecule is called the partial Wigner transform in 
Refs.~\cite{kapral_mixed_1999,kapral_progress_2006}.

Transforming (\ref{LDyson_GE}) to the basis of many-body adiabatic states of 
molecular electron subsystem, $\{\lvert \alpha,R \rangle\}$ leads to
(see Appendix~\ref{appB} for details)
\begin{align}
  \label{LDyson_AS}
    & i\frac{\partial}{\partial\tau} \mathbf{G}(\tau,\tau';R,P)
    = \mathbf{I}\,\delta(\tau,\tau') 
    \\ & +
    \left[
      -\frac{1}{2M}\left(\frac{1}{4}\frac{\partial^2}{\partial R^2}
      +iP\frac{\partial}{\partial R}-P^2\right)\mathbf{I}+\mathbf{E}(R)
      \right]\mathbf{G}(\tau,\tau';R,P)
   \nonumber \\ &
    -\frac{1}{8M}\bigg\{\mathbf{h}(R),\mathbf{G}(\tau,\tau';R,P)\bigg\}
    -\frac{i\, P}{2M}\bigg[\mathbf{d}(R),\mathbf{G}(\tau,\tau';R,P)\bigg]    
   \nonumber \\ &
    -\frac{1}{4M}\frac{\partial}{\partial R}\bigg[\mathbf{d}(R),\mathbf{G}(\tau,\tau';R,P)\bigg]
    -\frac{i}{2}\mathbf{F}(R)\,\frac{\partial}{\partial P} \mathbf{G}(\tau,\tau';R,P)
     \nonumber \\ & 
      +\frac{1}{4M}\mathbf{d}(R)\,\mathbf{G}(\tau,\tau';R,P)\,\mathbf{d}(R)
     \nonumber \\ &
      +\int_c d\tau''\,
      \mathbf{\Sigma}(\tau,\tau'';R,P)\,
      \mathbf{G}(\tau'',\tau';R,P)
      \nonumber
\end{align}
Here, $\mathbf{G}$, $\mathbf{h}$, $\mathbf{d}$, $\mathbf{F}$ are matrices in the electron subspace 
\begin{equation}
\label{GhdF_def}
  \begin{split}
  & G_{\alpha,\alpha'}(\tau,\tau';R,P) =
  \\ &
 \int dq\int dq'\,
    \Psi_{\alpha}^{*}(q,R) G(q,\tau;q',\tau'; R,P) \Psi_{\alpha'}(q',R)
  \\ &
    h_{\alpha\alpha'}(R)=\langle\alpha,R\rvert \frac{\partial^2}{\partial R^2}
    \lvert \alpha',R\rangle
  \\ &
    d_{\alpha\alpha'}(R)=\langle\alpha,R\rvert \frac{\partial}{\partial R}
    \lvert \alpha',R\rangle
  \\ &
    F_{\alpha\alpha'}(R)=\langle\alpha,R\rvert -\frac{\partial V}{\partial R}
    \lvert\alpha',R\rangle
  \end{split}
\end{equation}
$\mathbf{E}(R)$ is diagonal matrix of the adiabtic states eigenenergies,
$\mathbf{\Sigma}$, Eq.~\ref{appB_SE}, is the self-energy which accounts for
effect of the thermal bath and contact degrees of freedom on the molecule, and $\mathbf{I}$ is the identity matrix.
Note that expression (\ref{LDyson_AS}) is the left side (derivative in  $\tau$)
Dyson equation. Similarly, one can derive an expression for the right side
(derivative in $\tau'$) Dyson equation (see Appendix~\ref{appC}).


\section{Equation of motion for the generalized density matrix}\label{dm}
While expression (\ref{LDyson_AS}) is the general final result of our
derivation, to compare it to previous considerations, 
we have to derive the equation of motion for the system density matrix.
We note that the standard density matrix is (up to a constant) 
a lesser projection of the pseudoparticle Green's function taken at equal times
\begin{equation}
\label{rho_standard}
\rho_{\alpha\alpha'}(t;R,P) \equiv i\zeta_\alpha G^{<}_{\alpha\alpha'}(t,t;R,P)
\end{equation}
where $\zeta_\alpha=+1$ ($-1$) for Bose (Fermi) state $\lvert\alpha,R\rangle$.
Here, we introduce a generalized (energy-resolved) density matrix
\begin{equation}
  \label{rho_def}
  \rho_{\alpha\alpha'}(T,R\vert E,P) \equiv i\zeta_\alpha G^{<}_{\alpha\alpha'}(T,E;R,P)
\end{equation}
where
\begin{equation}
  G^{<}_{\alpha\alpha'}(T,E;R,P)=\int d(t-t')\, e^{iE(t-t')}
  G^{<}_{\alpha\alpha'}(T,t-t';R,P)
\end{equation}
with $T\equiv(t+t')/2$ and $G^{<}_{\alpha\alpha'}(T,t-t';R,P)$
is the Wigner representation (in time variables) of the lesser projection
$G^{<}_{\alpha\alpha'}(t,t';R,P)$.
The standard density matrix (\ref{rho_standard}) is obtained by integrating (\ref{rho_def}) in energy $E$.

Using lesser projections of equations (\ref{LDyson_AS}) and (\ref{appC_RDyson_AS})
yields equation-of-motion for the density matrix (\ref{rho_def})
(see Appendix~\ref{appD} for derivation)
\begin{equation}
  \label{rho_EOM}
  \begin{split}
    & \frac{\partial}{\partial T}\rho_{\alpha\alpha'}(T,R\vert E,P)
    =-i\sum_{\beta,\beta'}\int dE' \int dP'\,
    \\ & \quad
    \mathcal{L}_{\alpha\alpha';\beta\beta'}(T,R\vert E,P;E',P')\,
    \rho_{\beta\beta'}(T,R\vert E',P')
  \end{split}
\end{equation}
Here,
\begin{equation}
  \label{calL_def}
  \begin{split}
    & \mathcal{L}_{\alpha\alpha';\beta\beta'}(T,R\vert E,P; E',P') = 
    \\ & \qquad
    \delta_{\alpha,\beta}\delta_{\alpha',\beta'}\delta(E-E')\delta(P-P')
    L_{\alpha\alpha'}(T,R\vert E,P)
      \\ & \qquad
    +i\, J_{\alpha\alpha';\beta\beta'}(T,R\vert E,P; E',P')
 \end{split}
 \end{equation}
 with
 \begin{widetext}
 \begin{equation}
 \begin{split}
  & L_{\alpha\alpha'}(T,R\vert E,P) \equiv 
    \omega_{\alpha\alpha'}(R)
    -i\frac{P}{M}\frac{\partial}{\partial R}
    -\frac{i}{2}\bigg[F_{\alpha\alpha}(R)+F_{\alpha'\alpha'}(R)\bigg]
    \frac{\partial}{\partial P}
        +\Sigma^r_{\alpha\alpha}(T,E;R,P)-\Sigma^a_{\alpha'\alpha'}(T,E;R,P)
 \\&
    J_{\alpha\alpha';\beta\beta'}(T,R\vert E,P; E',P') = \delta(E-E')\delta(P-P')
    \bigg(
    -\frac{P}{M}\left[
      \delta_{\alpha',\beta'}\, d_{\alpha\beta}(R)
      \left(1+\frac{1}{2}S_{\alpha\beta}\frac{\partial}{\partial P}\right)
      +\delta_{\alpha,\beta}\, d_{\alpha'\beta'}^{*}
      \left(1+\frac{1}{2}S_{\alpha'\beta'}^{*}\frac{\partial}{\partial P}\right)
    \right]
        \\ & \quad
    - i\bigg[
      \delta_{\alpha',\beta'}\left(1-\delta_{\alpha,\beta}\right)
      \Sigma^r_{\alpha\beta}\left(T,E;R,P)\right)
      -\delta_{\alpha,\beta} \left(1-\delta_{\alpha',\beta'}\right)
      \Sigma^a_{\beta'\alpha'}\left(T,E;R,P)\right) 
      \bigg]
      \bigg)
          \\ & \quad
      -\zeta_\alpha\,\zeta_\beta\sum_\gamma\bigg[    
       G^{r}_{\alpha\gamma}(T,E;R,P)\,
       \eta^{<}_{\gamma\beta,\alpha'\beta'}(T,E-E';R,P-P')
        -\eta^{<}_{\alpha\beta,\gamma\beta'}(T,E-E';R,P-P')\, 
        G^{a}_{\gamma\alpha'}(T,E;R,P)
        \bigg]
    \end{split}
\end{equation}
\end{widetext}
where $\mathbf{\Sigma}^{r}$ and $\eta^{<}$ are defined, respectively, 
in (\ref{appD_SE}) and (\ref{appD_eta_def}),  
$\mathbf{\Sigma}^{a}\equiv\left[\mathbf{\Sigma}^r\right]^\dagger$,
and
\begin{equation}
\begin{split}
\omega_{\alpha\alpha'}(R) &\equiv E_\alpha(R)-E_{\alpha'}(R)
\\
S_{\alpha\beta}(R) &\equiv \left( F_{\alpha\beta}(R) - \delta_{\alpha,\beta} F_{\alpha\alpha}(R) \right)\left(\frac{P}{M}\, d_{\alpha\beta}(R)\right)^{-1}
\\ &
=\omega_{\alpha\beta}(R)\, d_{\alpha\beta}(R)\left(\frac{P}{M}\, d_{\alpha\beta}(R)\right)^{-1}
\end{split}
\end{equation}
Note that in the absence of a thermal environment and contacts, the consideration
reduces to isolated molecule and Eq.(\ref{rho_EOM}) reproduces
result of Ref.~\cite{kapral_mixed_1999} after integration over $E$.

The solution of Eq.~(\ref{rho_EOM}) can be written down in the form of an infinite series
\begin{equation}
\label{infinite_series}
  \begin{split}
    &\rho_{\alpha\alpha'}(T,R\vert E,P) =
    e^{-iL_{\alpha\alpha'}T}\rho_{\alpha\alpha'}(0,R\vert E,P)
    \\ &+
    \int\frac{dE_1}{2\pi}\int\frac{dP_1}{2\pi}
    \sum_{\beta,\beta'}\int_0^T dT'\,
    \\ & \qquad
     e^{-iL_{\alpha\alpha'}(T-T')}
    J_{\alpha\alpha';\beta\beta'}(T',R\vert E,P; E_1,P_1) 
    \\ & \qquad
        e^{-iL_{\beta\beta'}T'}\rho_{\beta\beta'}(0,R\vert E_1,P_1)
    \\ &+
    \int\frac{dE_1}{2\pi}\int\frac{dP_1}{2\pi}
    \int\frac{dE_2}{2\pi}\int\frac{dP_2}{2\pi}
    \\ &\quad
    \sum_{\beta,\beta',\gamma,\gamma'}\int_0^T dT'\int_0^{T'} dT''\,
    \\ &\qquad
    e^{-i L_{\alpha\alpha'}(T-T')} \,\,
    J_{\alpha\alpha';\beta\beta'}(T',R\vert E,P;E_1,P_1)
    \\ &\qquad
    e^{-iL_{\beta\beta'}(T'-T'')}
    J_{\beta\beta';\gamma\gamma'}(T'',R\vert ,E_1,P_1;E_2,P_2)
    \\ & \qquad
    e^{-i L_{\gamma\gamma'} T''}\rho_{\gamma\gamma'}(0,R\vert E_2,P_2)
    \\ & +\ldots
  \end{split}
\end{equation}
Following Ref.~\cite{kapral_mixed_1999}, we identify terms in the series with a sequence of trajectory segments, which are determined by
$e^{-i L_{\alpha\alpha'}T}$ involving evolution of
of two adiabatic states $\alpha$ and $\alpha'$.
The segments are intercepted with quantum transition events described
by $J_{\alpha\alpha';\beta\beta'}$. The latter can be expressed as 
a sequence of surface hopping trajectories. 
However, such a representation is formal and does not reflect the system's actual time evolution (see comments below).

The important differences from the consideration in Ref.~\cite{kapral_mixed_1999} are:
\begin{enumerate}
  \item Due to the presence of the baths, there are no well-defined adiabatic
    surfaces. Instead, one has to deal with bath-induced 
    energy and momentum distributions that resemble standard state
    broadening when system states hybridize with states of the baths.
  \item Evolution described by
    $e^{-i L_{\alpha\alpha'} T}$ includes bath-induced
    dissipation. As a result, while following Ref.~\cite{kapral_mixed_1999} one can formally introduce quantities $\tilde R_{t_1,\alpha\alpha'}\equiv e^{-iL_{\alpha\alpha'}(t-t_1)}R$ and $\tilde P_{t_1,\alpha\alpha'}\equiv e^{-iL_{\alpha\alpha'}(t-t_1)}P$ to build explicit surface hopping representation, they cannot be identified with coordinate and momentum at time $t_1<t$ because backward evolution in time is governed by a different Liouvillian operator~\cite{stenholm_time_2004}.
    This indicates that the whole idea of surface hopping is questionable in open systems.
  \item Due to the open character of the system, total energy $E$ and nuclear momentum $P$ are not conserved anymore. They may change during surface hopping
    events described by $J_{\alpha\alpha';\beta\beta'}$ when the hopping is due to coupling to a bath.
  \item The detailed balance of hopping events is preserved due to fulfillment of the KMS conditions by the Green's function formulation and due to consistency of the gradient expansion procedure. The density matrix is constrained by the terms describing couplings to the baths.
  \item In the absence of the baths and after integrating over the energy, expression (\ref{infinite_series}) reduces to the results of Reef.~\cite{kapral_mixed_1999}.  
\end{enumerate}


\section{Connection to the fewest switches surface hopping}\label{SectionFSSH}
To establish a connection with the fewest switches surface hopping, we follow Ref.~\cite{kapral_surface_2016} and shift to a Lagrangian frame of reference that moves with the nuclei along a single adiabatic surface $\gamma$
\begin{equation}
\begin{split}
\frac{d}{dT} R_\gamma(T) &= \frac{P_\gamma(T)}{M}
\\
\frac{d}{dT} P_\gamma(T) &= -\nabla_{R_\gamma(T)} E_\gamma\left(R_\gamma(T)\right)
\end{split}
\end{equation}
In this frame, the evolution is
\begin{widetext}
\begin{equation}
  \label{rho_EOM_Lagrange}
 \begin{split}
    & \frac{\partial}{\partial T}\rho_{\alpha\alpha'}\left(T,R_\gamma(T)\vert E,P_\gamma(T)\right)
    =
    \bigg[-i\omega_{\alpha\alpha'}\left(R_\gamma(T)\right)
    -\frac{1}{2}\bigg(\Delta F_{\alpha\alpha,\gamma\gamma}\left(R_\gamma(T)\right)+\Delta F_{\alpha'\alpha',\gamma\gamma}\left(R_\gamma(T)\right)\bigg) \frac{\partial}{\partial P_\gamma(T)}
   \\ & \qquad
    -i\bigg(\Delta\Sigma^{r}_{\alpha\alpha,\gamma\gamma}(T,E;R_\gamma(T),P_\gamma(T))
    - \Delta\Sigma^{a}_{\alpha'\alpha',\gamma\gamma}(T,E;R_\gamma(T),P_\gamma(T))
    \bigg)
    \bigg]
    \rho_{\alpha\alpha'}\left(T,R_\gamma(T)\vert E,P_\gamma(T)\right)
    \\ & \qquad
    +\sum_{\beta,\beta'}\int dE'\int dP'\,
    J_{\alpha\alpha'\vert\beta\beta'}\left(T,R_\gamma(T)\vert E,P_\gamma(T);E',P'\right)
    \rho_{\beta\beta'}\left(T,R_\gamma(T)\vert E',P'\right)
    \\ & \approx
      \bigg[-i\omega_{\alpha\alpha'}\left(R_\gamma(t)\right)
    -\frac{1}{2}\bigg(\Delta F_{\alpha\alpha,\gamma\gamma}\left(R_\gamma(T)\right)+\Delta F_{\alpha'\alpha',\gamma\gamma}\left(R_\gamma(T)\right)\bigg) \frac{\partial}{\partial P_\gamma(T)}
   \\ & \qquad
    -i\bigg(\Delta\Sigma^{r}_{\alpha\alpha,\gamma\gamma}(T,E;R_\gamma(T),P_\gamma(T))
    - \Delta\Sigma^{a}_{\alpha'\alpha',\gamma\gamma}(T,E;R_\gamma(T),P_\gamma(T))
    \bigg)
    \bigg]
    \rho_{\alpha\alpha'}\left(T,R_\gamma(T)\vert E;P_\gamma(T)\right)
    \\ & \qquad
    +\sum_{\beta,\beta',\gamma'}\int dE'\,
    J_{\alpha\alpha'\vert\beta\beta'}\left(T,R_\gamma(T)\vert E,P_\gamma(T);E',P_{\gamma'}(T)\right)
    \rho_{\beta\beta'}\left(T,R_\gamma(T)\vert E',P_{\gamma'}(T)\right)
\end{split}
\end{equation}
where
\begin{equation}
\begin{split}
 &\Delta F_{\alpha\alpha,\gamma\gamma}(R_\gamma(T)) \equiv
F_{\alpha\alpha}\left(R_\gamma(T)\right) -
F_{\gamma\gamma}\left(R_\gamma(T)\right)
 \\
&\Delta\Sigma^{r,a}_{\alpha\alpha,\gamma\gamma}(T,E;R_\gamma(T),P_\gamma(T))
\equiv
 \Sigma^{r,a}_{\alpha\alpha}(T,E;R_\gamma(T),P_\gamma(T))
- \Sigma^{r,a}_{\gamma\gamma}(T,E;R_\gamma(t),P_\gamma(T))
\end{split}
\end{equation}
Following Ref.~\cite{kapral_surface_2016}, we introduce the momentum jump approximation
\begin{equation}
\begin{split}
& d_{\alpha\beta}(R)\bigg(\frac{P}{M}+\frac{1}{2}S_{\alpha\beta}\frac{\partial}{\partial P}\bigg)
= \frac{P}{M}\, d_{\alpha\beta}(R)\bigg(1+M\,\omega_{\alpha\beta}(R)\,\frac{\partial}{\partial\mathcal{Y}_{\alpha\beta}(R)}\bigg)
\\ &
\approx \frac{P}{M}\, d_{\alpha\beta}(R)\,\, \exp\bigg[M\,\omega_{\alpha\beta}(R)\,\frac{\partial}{\partial \mathcal{Y}_{\alpha\beta}(R)}\bigg]
\equiv \frac{P}{M}\, d_{\alpha\beta}(R)\,\hat j_{\alpha\beta}
\end{split}
\end{equation}
where
$
\mathcal{Y}_{\alpha\beta} \equiv \bigg(P\cdot d_{\alpha\beta}(R)\bigg)^2
$.
Within the approximation, disregarding cross-state dissipation and bath-induced coherences between transitions, we get the $\gamma\alpha'$ density matrix element
\begin{equation}
\label{FSSH}
\begin{split}
&\frac{\partial}{\partial T}\rho_{\gamma\alpha'}\left(T,R_\gamma(T)\vert E,P_\gamma(T)\right) \approx
\\ &
\bigg[-i\omega_{\gamma\alpha'}\left(R_\gamma(T)\right)
-\frac{1}{2}\,\Delta F_{\alpha'\alpha',\gamma\gamma}\left(R_\gamma(T)\right)\,\frac{\partial}{\partial P_\gamma(T)}
+i\,\Delta\Sigma^{a}_{\alpha'\alpha',\gamma\gamma}\left(T,E;R_\gamma(T),P_\gamma(T)\right)
\bigg]\rho_{\gamma\alpha'}\left(T,R_\gamma(T)\vert E,P_\gamma(T)\right)
\\ &
-\sum_\beta\frac{P_\gamma(T)}{M}\bigg[ d_{\gamma\beta}(R_\gamma(T))\,\hat j_{\gamma\beta}\left(R_\gamma(T),P_\gamma(T)\right)\,\rho_{\beta\alpha'}\left(T,R_\gamma(T)\vert E,P_\gamma(T)\right)
\\ & \qquad\qquad\quad
- d^{*}_{\alpha'\beta}\left(R_\gamma(T)\right)\, \hat j^{*}_{\alpha'\beta}\left(R_\gamma(T),P_\gamma(T)\right)\,\rho_{\gamma\beta}\left(T,R_\gamma(T)\vert E,P_\gamma(T)\right)\bigg]
\\ &
- \sum_{\beta,\gamma'}\zeta_\gamma\zeta_\beta\int dE'\bigg[
G^r_{\gamma\gamma}\left(T,E;R_\gamma(T),P_\gamma(T)\right)
\delta_{\alpha',\gamma}\eta^{<}_{\gamma\beta,\gamma\beta}\left(T,E-E';R_{\gamma}(T),\Delta P_{\gamma\gamma'}(T)\right)
\\ &\qquad\qquad\qquad\quad\,\,\,
-\eta^{<}_{\gamma\beta,\gamma\beta}\left(T,E-E';R_\gamma(T),\Delta P_{\gamma\gamma'}(T)\right)
G^a_{\alpha'\alpha'}\left(T,E;R_\gamma(T),P_\gamma(T)\right)
\bigg]\rho_{\beta\beta}\left(T,R_\gamma(T)\vert E',P_{\gamma'}(T)\right)
\end{split}
\end{equation}
\end{widetext}
Here, $\Delta P_{\gamma\gamma'}(T)\equiv P_\gamma(T)-P_{\gamma'}(T)$.
Up to the terms associated with the self-energy (i.e., terms containing $\Delta\Sigma^a$ and $\eta^{<}$), this expression is equivalent to Eqs.~(23) (for $\alpha'=\gamma$) and (24) (for $\alpha'\neq\gamma$) of Ref.~\cite{kapral_surface_2016}. These equations are the starting point for the analysis of the approximations needed to derive the FSSH scheme. 
In principle, one could repeat this analysis for the generalized version of the equations presented in Eq.~(\ref{FSSH}), obtaining a generalized (top account for the presence of the baths) version of the transition rate to be employed within the FSSH.
However, even without performing the analysis, we can come to conclusions similar to those reached in Ref.~\cite{kapral_surface_2016}:
\begin{enumerate}
\item To derive Eq.(\ref{FSSH}) we had to disregard all inter-state and inter-transition coherences. While this assumption is a good approximation for well-separated adiabatic surfaces, it becomes unreasonable at the crossing point. As a result, the description of dissipation becomes significantly incorrect.
\item Local detailed balance is still satisfied even after the approximations used to reach Eq.(\ref{FSSH}). It is guaranteed by the presence of the baths, whose correlation functions (self-energies) satisfy the KMS conditions. In the case of an isolated molecule (no baths present), the local detailed balance (and, as a result, proper thermodynamic description of the system) is not guaranteed.
\item In the presence of the baths, energy resolution of the density matrix, Eq.(\ref{rho_def}), is crucial for proper energy conservation. The latter is enforced due to the presence of $E-E'$ dependence in $\eta^{<}$.
\end{enumerate}
Thus, the FSSH algorithm can hardly describe quantum nonadiabatic dynamics properly, and multiple efforts in {\em ad hoc} modifications are unlikely to fix its fundamental flaws.


\section{Conclusions}\label{conclusion}
Nonadiabatic molecular dynamics (NAMD) is central for the theoretical description of many processes, including charge and energy transfer, solar energy conversion, intersystem crossing, and chemistry. Most studies treat nuclear degrees of freedom classically, while treating the electron subsystem quantum mechanically.
The majority of these studies dealt with isolated molecular systems.
Efforts to generalize the formulation to open systems fall into three categories: a brute-force approach (treating the system and bath as one isolated supersystem), more elaborate ad hoc formulations, and a first-principles derivation of quantum-classical equations. The latter formulation extended the isolated-system consideration by separating classical and quantum dynamics, followed by tracing out parts of the (initially isolated) supersystem. 
Limitations of such a consideration are 1.~Impossibility of considering nonequilibrium situations with several baths (each at its own equilibrium) coupled to the molecule and 2.~Order of operators (transfer to classicality before tracing out bath degrees of freedom), which partially disregards quantum effects at the molecule-bath interface.

We present a first-principles derivation of quantum-classical evolution in open molecular systems. We employ a nonequilibrium pseudoparticle Green's function approach as a fully quantum formulation of dynamics for an open molecular system (treating both nuclear and electronic degrees of freedom quantum mechanically).
This formulation provides a starting point for considering a quantum-to-classical transition of nuclear dynamics in an open system and for deriving the equation of motion for a generalized (energy-resolved) density matrix. We compare our results with earlier formulations of mixed quantum-classical dynamics and analyze approximations necessary to reduce our consideration to the FSSH formulation.
Our main conclusions are
\begin{enumerate}
 \item In the presence of the baths, there are no well-defined adiabatic
    surfaces. Instead, one has to deal with bath-induced 
    energy distribution that resembles standard state
    broadening when system states hybridize with states of the baths.
    As a result, the whole idea of hopping between surfaces is questionable in open systems. 
  \item Due to the open character of the system, total energy $E$ and nuclear momentum $P$ are not conserved anymore. They may change during surface hopping
    events due to coupling to a bath. We note that energy resolution of the density matrix, Eq.(\ref{rho_def}), is crucial for proper account of energy conservation in the whole universe (system plus baths).
  \item The detailed balance of hopping events is preserved due to fulfillment of the KMS conditions by the bath Green's function formulation and due to consistency of the gradient expansion procedure. It is imposed on the density matrix by the terms describing couplings to the baths.
 In the case of an isolated molecule (no baths present), the local detailed balance (and, as a result, proper thermodynamic description of the system) is not guaranteed.
\item To derive the FSSH equations, we must disregard all inter-state and inter-transition coherences. While this assumption is a good approximation for well-separated adiabatic surfaces, it becomes unreasonable at the crossing point. As a result, the description of dissipation becomes significantly incorrect. Thus, the FSSH algorithm can hardly describe quantum nonadiabatic dynamics properly, and multiple efforts in {\em ad hoc} modifications are unlikely to fix its fundamental flaws.
\end{enumerate}

Finally, we note that traditional treatments of open systems using methods originally developed for isolated molecules are minimally consistent and maximally questionable. For example, theoretical descriptions of photo- and bias-induced chemistry for molecules on surfaces and at interfaces are similarly questionable.
Indeed, the very concepts of adiabatic surface, conical intersection, etc. are defined only for an isolated molecule. Extending these formulations to open, nonequilibrium molecular systems is a direction for future research.


\appendix

\section{Derivation of the Dyson equation (\ref{Dyson})}\label{appA}
Here, starting from a full quantum mechanical description of the universe, we derive the Dyson equation (\ref{Dyson}) and provide explicit expressions for the self-energies $\Pi$ and $\Sigma$.

We start by rewriting the Hamiltonian (\ref{H}) in second quantization. 
To do this, we separate nuclear states into system and thermal-bath parts, assuming the latter is described as a continuum of non-interacting modes at thermal equilibrium. 
We make a similar assumption for the electronic states of the bath.
Let us represent many-body molecular eigenstates $\Psi_S(q, Q)$ employing pseudoparticles $\hat p_S$, and single-particle degrees of freedom in the baths, $\phi_\alpha(Q)$ and $\psi_k(q)$, using standard second quantization by introducing annihilation operators $\hat a_\alpha$ and $\hat c_k$, respectively. 
This leads to
\begin{equation}
\label{appA_HSQ}
\begin{split}
\hat H_{SQ} &= \sum_S E_{S} \hat p_S^\dagger\hat p_S
+ \sum_\alpha\omega_\alpha a^\dagger_\alpha a_\alpha
+ \sum_k \varepsilon_k \hat c_k^\dagger\hat c_k
\\ &
+ \sum_{S_1,S_2,\alpha} \bigg(U_{S_1S_2,\alpha}\hat p_{S_2}^\dagger\hat p_{S_1}\hat a_\alpha + H.c.\bigg) 
\\ &
+\sum_{S_1,S_2,k}\bigg(V_{S_1S_2,k}\hat p_{S_2}^\dagger\hat p_{S_1}\hat c_k + H.c.\bigg) 
\end{split}
\end{equation}
Here, $E_S$ are many-body eigenenergies of the molecular subsystem,
\begin{equation}
\begin{split}
U_{S_1S_2,\alpha} &\equiv \int dq\int dQ\, \Psi^{*}_{S_2}(q,Q)\,\hat H\,\Psi_{S_1}(q,Q)\phi_\alpha(Q)
\\
V_{S_1S_2,k} &\equiv \int dq\int dQ\, \Psi^{*}_{S_2}(q,Q)\,\hat H\,\Psi_{S_1}(q,Q)\psi_k(q)
\end{split}
\end{equation}

We are interested in the pseudoparticle Green's function
\begin{equation}
\label{appA_G}
G_{SS'}(\tau,\tau') \equiv -i\langle T_c\,\hat p_{S}(\tau)\,\hat p_{S'}(\tau')\rangle
\end{equation}
Following the standard formulation, one can derive the Dyson equation for the Green's function
\begin{equation}
\label{appA_Dyson}
\begin{split}
& i\frac{\partial}{\partial\tau} G_{SS'}(\tau,\tau') = E_S G_{SS'}(\tau,\tau')
\\ &
+\sum_{S_1}\int_c d\tau_1\, \Sigma_{SS_1}(\tau,\tau_1)\,
 G_{S_1S'}(\tau_1,\tau')
\end{split}
\end{equation}
The many-body character of the molecule-bath couplings does not allow derivation of exact expressions for the self-energies. Within the lowest (second) order, the approximate expressions are
\begin{equation}
\label{appA_SENCA}
\begin{split}
\Sigma_{SS'}(\tau,\tau') &= i\sum_{S_1,S_2} G_{S_1S_2}(\tau,\tau')
\\ &\times
\bigg[\pi_{SS_1,S'S_2}(\tau,\tau')+\pi_{S_2S',S_1S}(\tau',\tau)
\\ & \,\,\,
+\sigma_{SS_1,S'S_2}(\tau,\tau')-\sigma_{S_2S',S_1S}(\tau',\tau)\bigg]
\end{split}
\end{equation}
where
\begin{equation}
\begin{split}
\pi_{SS_1,S'S_2}(\tau,\tau') &=
\sum_\alpha U_{SS_1,\alpha}\, U_{\alpha,S'S_2}\, 
\\ &\times
(-i)\langle T_c\, \hat a_\alpha(\tau)\,\hat a_\alpha^\dagger(\tau')\rangle_0
\\
\sigma_{SS_1,S'S_2}(\tau,\tau') &=
\sum_k V_{SS_1,k}\, V_{k,S'S_2}\, 
\\ &\times
(-i)\langle T_c\, \hat c_k(\tau)\,\hat c_k^\dagger(\tau')\rangle_0
\end{split}
\end{equation}
Subscript $0$ indicates free evolution (i.e., evolution in the baths in the absence of baths-molecule couplings).

Molecular quantum fields $\hat\Psi(q, Q)$ are expressed in terms of the pseudoparticle operators via expansion in the basis of molecular many-body states $\Psi_S(q, Q)$
\begin{equation}
\label{appA_transform}
\hat\Psi(q, Q) = \sum_S \hat p_S\,\Psi_S(q, Q)
\end{equation}
Green's function (\ref{G}) is obtained by applying the transformation to (\ref{appA_G})
Similarly, the Dyson equation (\ref{Dyson}) is given by the basis transformation applied to (\ref{appA_Dyson}), and its self-energy (\ref{appA_SE}) reads
\begin{equation}
\label{appA_SE}
\begin{split}
&\Sigma(X,\tau;X',\tau') = i\int dX_1\int dX_2\, G(X_1,\tau;X_2,\tau')
\\ &\times
\bigg[\pi(X,X_1,\tau;X',X_2,\tau')+\pi(X_2,X',\tau';X_1,X,\tau)
\\ & \,\,\,
+\sigma(X,X_1,\tau;X',X_2,\tau')-\sigma(X_2,X',\tau';X_1,X,\tau)\bigg]
\end{split}
\end{equation}
where we introduced shorthand notation $X\equiv q,Q$.
 
\section{Dyson equation in the basis of adiabatic states}\label{appB}
To derive the Dyson equation in the basis of adiabatic states, we must evaluate 
\begin{equation}
\int dq\int dq' \Psi_\alpha^{*}(q,R)\ldots \Psi_{\alpha'}(q',R)
\end{equation}
 matrix elements of terms in Eq.~(\ref{LDyson_GE}). Here,
 $\Psi_{\alpha}(q,R)$ are adiabatic states; that is,
 \begin{equation}
 \label{appB_AS}
 \hat h(q,R)\, \Psi_{\alpha}(q,R) = E_\alpha(R)\, \Psi_{\alpha}(q,R)
\end{equation}
The following relations are useful
\begin{equation}
\label{appB_relations}
\begin{split}
&
\int dq\int dq'\, \Psi_\alpha^{*}(q,R)\,G(q,q';R,P;\tau\tau')\, \Psi_{\alpha'}(q',R)
\\ & \qquad
= G_{\alpha\alpha'}(R,P;\tau,\tau')
\\ &
\int dq\int dq'\, \Psi_\alpha^{*}(q,R)\, \frac{\partial}{\partial R} G(q,q';R,P;\tau,\tau')\, \Psi_{\alpha'}(q',R) 
\\ & \qquad = 
\frac{\partial}{\partial R}G_{\alpha\alpha'}(R,P;\tau,\tau')
\\ & \qquad
-\sum_{\beta}\bigg(G_{\alpha\beta}(R,P;\tau,\tau')\, d_{\beta\alpha'}(R)
\\ & \qquad\qquad
- d_{\alpha\beta}(R)\, G_{\beta\alpha'}(R,P;\tau,\tau')\bigg)
\\ &
\int dq\int dq'\, \Psi_\alpha^{*}(q,R)\, \frac{\partial^2}{\partial R^2} G(q,q';R,P;\tau\tau')\, \Psi_{\alpha'}(q',R) 
\\ & \qquad = 
\frac{\partial^2}{\partial R^2} G_{\alpha\alpha'}(R,P;\tau,\tau')
\\ &
+\sum_{\beta}\big(F_{\alpha\beta}(R)\, G_{\beta\alpha'}(R,P;\tau,\tau')
\\ & \qquad\qquad
+G_{\alpha\beta}(R,P;\tau,\tau')\, F_{\beta\alpha'}(R,P;\tau,\tau')\bigg)
\\ & \qquad
+2\frac{\partial}{\partial R}\sum_{\beta}\bigg(d_{\alpha\beta}(R)\, G_{\beta\alpha'}(R,P;\tau,\tau')
\\ & \qquad\qquad
-G_{\alpha\beta}(R,P;\tau,\tau')\, d_{\beta\alpha'}(R)\bigg)
\\ & \qquad
-2\sum_{\beta,\beta'}d_{\alpha\beta}(R)\, G_{\beta\beta'}(R,P;\tau,\tau')\, d_{\beta'\alpha'}(R)
\end{split}
\end{equation}
where $d_{\alpha\alpha'}(R)$ and $F_{\alpha\alpha'}(R)$ are defined in (\ref{GhdF_def}).
Employing (\ref{appB_AS}) and (\ref{appB_relations}) in (\ref{LDyson_GE}) leads to (\ref{LDyson_AS}).

We now derive an expression for the self-energy $\mathbf{\Sigma}(\tau,\tau'; R, P)$.
Starting from (\ref{appA_SE}) and making an assumption of slow nuclear motion compared to the bath dynamics,
\begin{equation}
\begin{split}
&\pi(X_1,X_2,\tau;X_3,X_4,\tau') \approx \delta(Q_1-Q_2)\,\delta(Q_3-Q_4)\,
\\ &\qquad\times
\pi(q_1,q_2,Q_1,\tau;q_3,q_4,Q_3,\tau'),
\\
&\sigma(X_1,X_2,\tau;X_3,X_4,\tau') \approx \delta(Q_1-Q_2)\,\delta(Q_3-Q_4)\,
\\ &\qquad\times
\sigma(q_1,q_2,Q_1,\tau;q_3,q_4,Q_3,\tau'),
\end{split}
\end{equation}
we get
\begin{equation}
\begin{split}
&\Sigma(q,\tau;q',\tau';R,P) =
\\ &\qquad
 i\int dq_1\int dq_2\int dP'\, 
G(q_1,\tau;q_2,\tau';R,P')
\\ &\qquad
\bigg[\,\,\,\,
  \pi(q,q_1,\tau;q',q_2,\tau';R,P-P') 
  \\ &\qquad
  + \pi(q_2,q',\tau';q_1,q,\tau;R,P'-P)
  \\ &\qquad
 +\sigma(q,q_1,\tau;q',q_2,\tau';R,P-P')
 \\ &\qquad
  - \sigma(q_2,q',\tau';q_1,q,\tau;R,P'-P)
\bigg]
\end{split}
\end{equation}
Finally, transforming to the basis of many-body adiabatic states of the electron subsystem, $\{\lvert\alpha, R\rangle\}$, leads to
\begin{align}
\label{appB_SE}
&\Sigma_{\alpha\alpha'}(\tau,\tau';R,P) = i\sum_{\beta,\beta'}\int dP'\, 
G_{\beta\beta'}(\tau,\tau';R,P')
\\ &
\bigg[\,\,\,\,
  \pi_{\alpha\beta,\alpha'\beta'}(\tau,\tau';R,P-P') + \pi_{\beta'\alpha',\beta\alpha}(\tau',\tau;R,P'-P)
 \nonumber \\ &
 +\sigma_{\alpha\beta,\alpha'\beta'}(\tau,\tau';R,P-P') - \sigma_{\beta'\alpha',\beta\alpha}(\tau',\tau;R,P'-P)
\bigg]
\nonumber
\end{align}

\begin{widetext}
\section{Right-side Dyson equation in the basis of adiabatic states}\label{appC}
Similar to (\ref{LDyson_AS}), one can derive the right-side version of the equation.
The derivation starts from the right-side analog of (\ref{Dyson})
\begin{equation}
\label{appC_RDyson}
\begin{split}
   &-i\frac{\partial}{\partial\tau'} G(q,Q,\tau;q',Q',\tau') =
  \left[-\frac{1}{2M}\frac{\partial^2}{\partial {Q'}{}^2}
  -\frac{1}{2m}\frac{\partial^2}{\partial {q'}{}^2}+V(q',Q')\right]
  G(q,Q,\tau;q',Q',\tau')
  \\ & \qquad
    + \int_c d\tau'' \int dq''\int dQ''\, G(q,Q,\tau;q'',Q'',\tau'')
    \Sigma(q'',Q'',\tau'';q',Q',\tau')
        +\delta(q-q')\delta(Q-Q')\delta(\tau,\tau')  
\end{split}
\end{equation}
which, after gradient expansion in the nuclear coordinates, yields the analog of (\ref{LDyson_GE})
\begin{align}
  \label{appC_RDyson_GE}
    & -i\frac{\partial}{\partial\tau'} G(q,\tau;q',\tau';R,P) =  
    \bigg[
      -\frac{1}{2M}\left(\frac{1}{4}\frac{\partial^2}{\partial R^2}
      -iP\frac{\partial}{\partial R} - P^2\right)
      + \hat h(q,R)
      -\frac{i}{2}\frac{\partial V(q,R)}{\partial R}\frac{\partial}{\partial P}
    \bigg] G(q,\tau;q',\tau';R,P)
  \nonumber \\ & \qquad
    +\int_c d\tau''\int dq''\, G(q,\tau;q'',\tau'';R,P)\,
    \Sigma(q'',\tau'';q',\tau';R,P)
    +\delta(q-q')\,\delta(\tau,\tau')
\end{align}
Evaluating adiabatic states matrix elements of the latter leads to
\begin{equation}
  \label{appC_RDyson_AS}
  \begin{split}
    & -i\frac{\partial}{\partial\tau'} \mathbf{G}(\tau,\tau';R,P)
    = \mathbf{I}\,\delta(\tau,\tau') +
    \mathbf{G}(\tau,\tau';R,P)\left[
      -\frac{1}{2M}\left(\frac{1}{4}\frac{\overset{\leftarrow}{\partial}{}^2}{\partial R^2}
      -iP\frac{\overset{\leftarrow}{\partial}}{\partial R}-P^2\right)\mathbf{I}+\mathbf{E}(R)
      \right]
    \\ & \qquad
    -\frac{1}{8M}\bigg\{\mathbf{h}(R),\mathbf{G}(\tau,\tau';R,P)\bigg\}
    +\frac{i\, P}{2M}\bigg[\mathbf{d}(R),\mathbf{G}(\tau,\tau';R,P)\bigg]    
    -\frac{1}{4M}\frac{\partial}{\partial R}\bigg[\mathbf{d}(R),\mathbf{G}(\tau,\tau';R,P)\bigg]
  \\ &    \qquad
  +\frac{1}{4M}\mathbf{d}(R)\,\mathbf{G}(\tau,\tau';R,P)\,\mathbf{d}(R)
    +\frac{i}{2}\frac{\partial}{\partial P} \bigg(\mathbf{G}(\tau,\tau';R,P)\bigg)\,\mathbf{F}(R) 
      +\int_c d\tau''\,
      \mathbf{G}(\tau,\tau'';R,P)\,
      \mathbf{\Sigma}(\tau'',\tau';R,P)
\end{split}
\end{equation}
which is the right-side version of (\ref{LDyson_AS}).

\section{Equation of motion for generalized density matrix}\label{appD}
We start by taking lesser projections of the left- and right-side Dyson equations, Eqs.~(\ref{LDyson_AS}) and (\ref{appC_RDyson_AS})
\begin{align}
  \label{appD_LDyson_lt}
    & +i\frac{\partial}{\partial t} \mathbf{G}^{<}(t,t';R,P)
    = 
    \left[
      -\frac{1}{2M}\left(\frac{1}{4}\frac{\partial^2}{\partial R^2}
      +iP\frac{\partial}{\partial R}-P^2\right)\mathbf{I}+\mathbf{E}(R)
      \right]\mathbf{G}^{<}(t,t';R,P)
   \\ & \qquad
    -\frac{1}{8M}\bigg\{\mathbf{h}(R),\mathbf{G}^{<}(t,t';R,P)\bigg\}
    -\frac{i\, P}{2M}\bigg[\mathbf{d}(R),\mathbf{G}^{<}(t,t';R,P)\bigg]  
    +\frac{1}{4M}\mathbf{d}(R)\,\mathbf{G}^{<}(t,t';R,P)\,\mathbf{d}(R)  
    \nonumber \\ & \qquad
    -\frac{1}{4M}\frac{\partial}{\partial R}\bigg[\mathbf{d}(R),\mathbf{G}^{<}(t,t';R,P)\bigg]
    -\frac{i}{2}\mathbf{F}(R)\,\frac{\partial}{\partial P} \mathbf{G}^{<}(t,t';R,P)
    \nonumber \\ & \qquad
      +\int dt''\bigg(
      \mathbf{\Sigma}^r(t,t'';R,P)\,\mathbf{G}^{<}(t'',t';R,P)
      + \mathbf{\Sigma}^{<}(t,t'';R,P)\,\mathbf{G}^a(t'',t';R,P)
      \bigg)
      \nonumber
\\  
\label{appD_RDyson_lt}
 & -i\frac{\partial}{\partial t'} \mathbf{G}^{<}(t,t';R,P)
    = 
    \mathbf{G}^{<}(t,t';R,P)
    \left[
      -\frac{1}{2M}\left(\frac{1}{4}\frac{\overset{\leftarrow}{\partial}{}^2}{\partial R^2}
      -iP\frac{\overset{\leftarrow}{\partial}}{\partial R}-P^2\right)\mathbf{I}+\mathbf{E}(R)
      \right]
 \\ & \qquad
    -\frac{1}{8M}\bigg\{\mathbf{h}(R),\mathbf{G}^{<}(t,t';R,P)\bigg\}
    +\frac{i\, P}{2M}\bigg[\mathbf{d}(R),\mathbf{G}^{<}(t,t';R,P)\bigg]    
    +\frac{1}{4M}\mathbf{d}(R)\,\mathbf{G}^{<}(t,t';R,P)\,\mathbf{d}(R)
  \nonumber \\ & \qquad
    -\frac{1}{4M}\frac{\partial}{\partial R}\bigg[\mathbf{d}(R),\mathbf{G}^{<}(t,t';R,P)\bigg]
    +\frac{i}{2}\frac{\partial}{\partial P} \bigg(\mathbf{G}(\tau,\tau';R,P)\bigg)\,\mathbf{F}(R)
      \nonumber \\ & \qquad
      +\int dt''\bigg(
      \mathbf{G}^{<}(t,t'';R,P)\,\mathbf{\Sigma}^a(t'',t';R,P)
      + \mathbf{G}^r(t,t'';R,P)\,\mathbf{\Sigma}^{<}(t'',t';R,P)
      \bigg)
      \nonumber
\end{align}
Next we subtract (\ref{appD_RDyson_lt}) from (\ref{appD_LDyson_lt}), perform Wigner transformation in time variable
\begin{equation}
T\equiv\frac{t+t'}{2} \qquad\mbox{and}\qquad s\equiv t-t',
\end{equation}
and Fourier transform in the quantum time $s\to E$ employing a gradient expansion in the self-energy term. This leads to
\begin{equation}
\label{appD_GltT}
\begin{split}
     i\frac{\partial}{\partial T} \mathbf{G}^{<}(T,E;R,P)
    &= 
    \bigg[\mathbf{E}(R),\mathbf{G}^{<}(T,E;R,P)\bigg]
    -\frac{iP}{M}\frac{\partial}{\partial R}\mathbf{G}^{<}(T,E;R,P)
    \\ &
    -\frac{i\, P}{M}\bigg[\mathbf{d}(R),\mathbf{G}^{<}(T,E;R,P)\bigg]   
    -\frac{i}{2}\bigg\{\mathbf{F}(R),\frac{\partial}{\partial P} \mathbf{G}^{<}(T,E;R,P)\bigg\}
     \\ &
      + \mathbf{\Sigma}^r(T,E;R,P)\,\mathbf{G}^{<}(T,E;R,P)
      - \mathbf{G}^{<}(T,E;R,P)\, \mathbf{\Sigma}^a(T,E;R,P)
      \\ &
      + \mathbf{\Sigma}^{<}(T,E;R,P)\,\mathbf{G}^a(T,E;R,P)
      - \mathbf{G}^r(T,E;R,P)\, \mathbf{\Sigma}^{<}(T,E;R,P)
\end{split}
\end{equation}

From (\ref{appB_SE}) we get~\cite{oh_transport_2011}
\begin{equation}
\label{appD_SE}
\begin{split}
&\Sigma^{r}_{\alpha\alpha'}(T,E;R,P) = 
i\sum_{\beta,\beta'}\int dE'\int dP'\, 
G^{r}_{\beta\beta'}(T,E';R,P')\, \eta_{\alpha\beta,\alpha'\beta'}^{>}(T,E-E';R,P-P')
\\ 
&\Sigma^{<}_{\alpha\alpha'}(T,E;R,P) =
i\sum_{\beta,\beta'}\int dE'\int dP'\, 
G^{<}_{\beta\beta'}(T,E';R,P')\, \eta_{\alpha\beta,\alpha'\beta'}^{<}(T,E-E';R,P-P')
\end{split}
\end{equation}
where
\begin{equation}
\label{appD_eta_def}
\begin{split}
\eta_{\alpha\beta,\alpha'\beta'}^{\gtrless}(T,E;R,P) &\equiv
\pi_{\alpha\beta,\alpha'\beta'}^{\gtrless}(T,E;R,P) + \pi_{\beta'\alpha',\beta\alpha}^{\lessgtr}(T,-E;R,-P)
\\ &
 +\,\sigma_{\alpha\beta,\alpha'\beta'}^{\gtrless}(T,E;R,P) - \sigma_{\beta'\alpha',\beta\alpha}^{\lessgtr}(T,-E;R,-P)
 \end{split}
\end{equation}
Substituting (\ref{appD_SE}) into (\ref{appD_GltT}) and employing definition (\ref{rho_def}) leads to (\ref{rho_EOM}).
\end{widetext}


%

\end{document}